\documentclass[twocolumn]{aastex62}
\defcitealias{2016kounkela}{Paper I}

\begin{document}
\title{VLBA Observations of Strong Anisotripic Radio Scattering towards the Orion Nebula}

\author[0000-0002-5365-1267]{Marina Kounkel}
\affil{Department of Physics and Astronomy, Western Washington University, 516 High St, Bellingham, WA 98225}
\author[0000-0003-1430-8519]{Lee Hartmann}
\affil{Department of Astronomy, University of Michigan, 1085 S. University st., Ann Arbor, MI 48109, USA}
\author[0000-0002-5635-3345]{Laurent Loinard}
\affil{Instituto de Radioastronom\'{i}a y Astrof\'{i}sica, Universidad Nacional Aut\'{o}noma de M\'{e}xico, Morelia 58089, Mexico}
\affil{Instituto de Astronom\'{i}a, Universidad Nacional Aut\'{o}noma de M\'{e}xico, Apartado Postal 70-264, CdMx 04510, Mexico}
\author{Amy J. Mioduszewski}
\affil{National Radio Astronomy Observatory, Domenici Science Operations Center, 1003 Lopezville Road, Socorro, NM 87801, USA}
\author[0000-0003-2737-5681]{Luis F. Rodr\'{i}guez}
\affil{Instituto de Radioastronom\'{i}a y Astrof\'{i}sica, Universidad Nacional Aut\'{o}noma de M\'{e}xico, Morelia 58089, Mexico}
\author[0000-0002-2863-676X]{Gisela N. Ortiz-Le\'{o}n}
\affil{Max Planck Institut f\"{u}̈r Radioastronomie, Auf dem H\"{u}̈gel 69,D-53121 Bonn, Germany}
\author[0000-0002-4120-3029]{Michael D. Johnson}
\affil{Harvard-Smithsonian Center for Astrophysics, 60 Garden Street, Cambridge, MA 02138, USA}
\author{Rosa M. Torres}
\affil{Centro Universitario de Tonal\'{a}, Universidad de Guadalajara, Avenida Nuevo Periférico No. 555, Ejido San Jos\'{e}, Tatepozco, C.P. 48525, Tonal\'{a}, Jalisco, M\'{e}xico}
\author[0000-0001-7124-4094]{Cesar Brice\~{n}o}
\affil{Cerro Tololo Interamerican Observatory, Casilla 603, La Serena, Chile}

\email{Marina.Kounkel@wwu.edu}

\begin{abstract}
We present observations of VLBA 20, a radio source found towards the edge of the Orion Nebula Cluster (ONC). Nonthermal emission dominates the spectral energy distribution of this object from radio to mid-infrared regime, suggesting that VLBA 20 is extragalactic. This source is heavily scattered in the radio regime. Very Long Baseline Array observations resolve it to $\sim 34\times19$ mas at 5 GHz, and the wavelength dependence of the scattering disk is consistent with $\nu^{-2}$ at other frequencies. The origin of the scattering is most likely the ionized X-ray emitting gas from the winds of the most massive stars of the ONC. The scattering is highly anisotropic, with the axis ratio of 2:1, higher than what is typically observed towards other sources.
\end{abstract}

\section{Introduction}

Interstellar scattering of extragalactic sources in the radio regime is a phenomenon caused by turbulent ionized gas located along the line of sight. It has been observed in a number of sources located behind Galactic HII regions and typically affects sub-GHz frequencies observations; comparatively, few sources are known in which scattering is observed at frequencies above a few GHz. The strongest known scatter broadening is found towards NGC 6334B, where the scattering disk has a size of 300 mas at 5 GHz \citep{trotter1998}. Other regions with strongly scattered sources include the galactic center \citep[e.g.,][]{backer1978,bower1998,ortiz-leon2016}, the Cygnus regions \citep[e.g.][]{wilkinson1994,desai2001}, B1849+005 \citep{lazio2004} and possibly M17 \citep{rodriguez2014}. Recently, \citet{kounkel2017} have also identified a region of scattering associated with the ionized gas region produced by a supernova in the $\lambda$ Ori cluster towards several sources, most strongly towards J0536+0944. In most of these cases, the scattering disk appears to be somewhat anisotropic, with the typical axial ratio of 1.2--1.5. This anisotropy is thought to be associated with the magnetic activity within the HII regions, with the preferential direction of the scattering aligned perpendicularly to the magnetic fields \citep{narayan1989}.

\begin{figure*}
 \centering 
        		\gridline{
        		  \fig{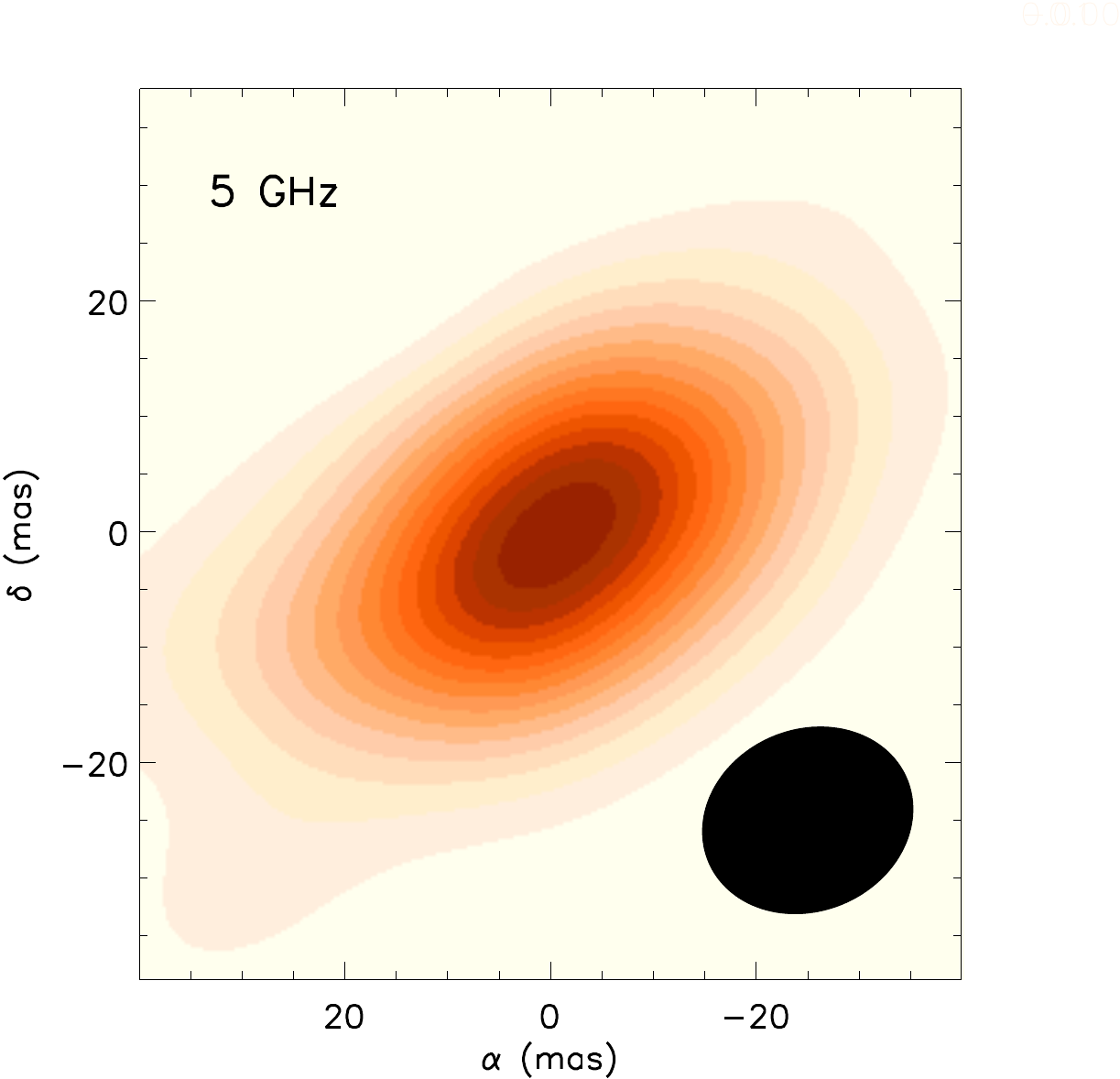}{0.35\textwidth}{}
			  \fig{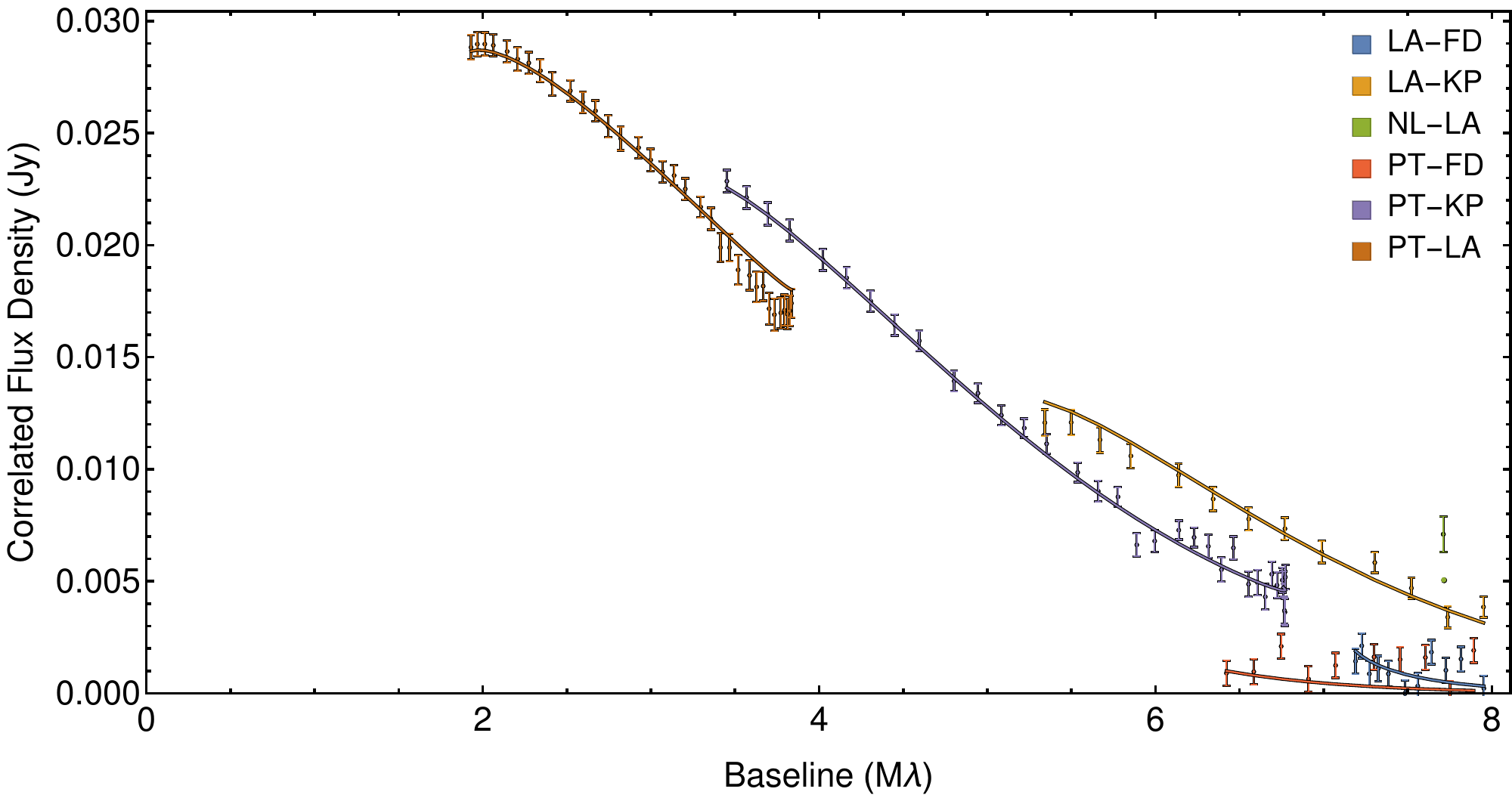}{0.6\textwidth}{}              
        }   
        
 \caption{VLBA 20, as seen at 5 GHz with VLBA, with data combined from 5 epochs observed. Left: Imaging of the source, with the UV range limited to 8 M$\lambda$, as no coherent emission is apparent beyond it. The corresponding synthesized beam is shown in black. Right: Visibility amplitude as a function of UV distance.\label{fig:scatter1} }
\end{figure*}

In the process of mapping the Orion Molecular Cloud Complex with the Very Long Baseline Array (VLBA), \citet{kounkel2017} have identified a peculiar source, VLBA 20 (Figure \ref{fig:scatter1}), which is located towards the west of the Orion Nebula Cluster (ONC). This source appeared to be resolved, and highly anisotropic, with a size of 34 $\times$ 19 mas at 5 GHz. No coherent emission was apparent at baselines beyond 8 M$\lambda$. Five epochs were observed in total, the morphology was largely consistent in all of them. In this paper we examine the nature of VLBA 20 and the origin of its morphology.

\section{Multi-wavelength observations}

\begin{figure}
\epsscale{1}
\plotone{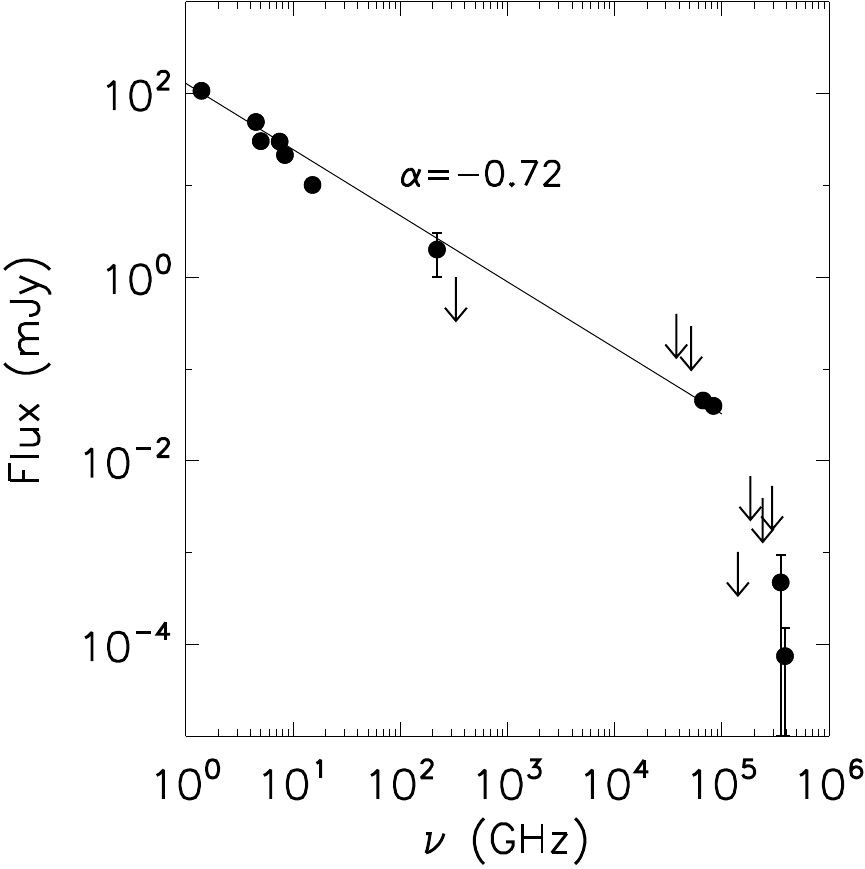}
\caption{Spectral energy distribution of VLBA 20, combining observations from VLA, VLBA, SMA, \textit{Spitzer}, and \textit{HST}. \label{fig:sed}}
\end{figure}

\begin{deluxetable}{cc}
\tabletypesize{\scriptsize}
\tablewidth{0pt}
\tablecaption{Fluxes of multi-wavelength observations of VLBA 20. \label{tab:flux}}
\tablehead{
\colhead{$\nu$} &\colhead{$F$}
}
\startdata
1.4 GHz & 106$\pm1$ mJy\\
4.5 GHz & 47.31$\pm2.37$ mJy\\
7.5 GHz & 32.95$\pm1.66$ mJy\\
224.521 GHz & 2 $\pm$ 1 mJy\\
336.425 GHz & $<1$ mJy\\
4.5 $\mu$m & $16.49 \pm 0.14$ mag\\
3.6 $\mu$m & $17.11\pm0.11$ mag\\
2.2 $\mu$m & $>$22 mag\\
1.6 $\mu$m & $>$20.9 mag\\
1.25 $\mu$m & $>$21.5 mag\\
1.03 $\mu$m & $>$22.1 mag\\
0.85 $\mu$m & $\sim$24 mag\\
0.775 $\mu$m & $\sim$27 mag\\
5 keV & $<6\times10^{-15}$ erg sec$^{-1}$cm$^{-2}$\\
\enddata
\end{deluxetable}

VLBA 20 (= GBS-VLA J053424.63-052838.5) was first identified in the Very Large Array (VLA) survey of the ONC \citep{kounkel2014}. It was observed as a point source at 4.5 and 7.5 GHz, with fluxes of $F_{4.5 GHz}=47.31\pm2.37$ mJy and $F_{7.5 GHz}=32.95\pm1.66$ mJy. Three different epochs have been obtained over a period of 2 months; no variability in flux was observed. Aside from \citet{kounkel2014} and \citet{kounkel2017}, no fluxes for VLBA 20 have been published. This source was originally identified as a young stellar object candidate; however, subsequent astrometric observations with VLBA have ruled out this interpretation due to a lack of any parallax or proper motions. In order to confirm that the detected radio emission is non-thermal as well as to determine the true nature of this source, we examined the archival data to construct its spectral energy distribution (SED). When possible, we also analyzed the variability of the source.

While not previously reported, a number of radio detections of VLBA 20 exist. Yusef-Zadeh has observed it with the VLA in four different configurations at 4.8 GHz between 1988 and 1989 (project code: AY24). We reduced and imaged all the archival data following the standard prescription in AIPS. The measured flux was $F_{4.8 GHz}=48\pm1$ mJy in all four epochs, consistent with our recent observations, showing no variability over a 25 year baseline. Additionally, it has been detected at 1.4 GHz in 1984 (Dulk, project code AD114) and 1987 (Yusef-Zadeh, project code AC170) with flux of $F_{1.4 GHz}=106\pm1$ mJy. In all the VLA observations, this source appears pointlike.

Only hints of the detection are available in any wavelength regimes other than radio. VLBA 20 has been previously detected with the \textit{Hubble} Space Telescope (HST) at 0.775 and 0.85 $\mu$m \citep{da-rio2009} and with the \textit{Spitzer} Space Telescope at 3.6 and 4.5 $\mu$m \citep{morales-calderon2011} during optical and infrared surveys of the Orion region. Unfortunately, due to the faintness of the source and insufficiently deep exposures, the HST detections are only barely significant and it is impossible to obtain reliable photometry. We estimate $m_{0.775 \mu m}\sim27$, $m_{0.85 \mu m}\sim25$. \textit{Spitzer} data yielded $m_{3.6\mu m} = 17.11\pm0.11$, $m_{4.5\mu m} = 16.49 \pm 0.14$. No sufficiently deep J, H, and K band exposures exist, but we estimate $m_Y > 22.1$, $m_J > 21.5$, $m_H > 20.9$ based on the sensitivity of the Galactic Cluster Survey performed as part of the United Kingdom Infrared Telescope (UKIRT) Infrared Deep Sky Survey, $m_K > 22$ based on the National Optical Astronomy Observatory Extremely Wide-Field Infrared Imager (NEWFIRM) observations towards that region (Megeath et al. in prep), as well as $m_{5.8\mu m} > 14$ and $m_{8\mu m} > 13$ from the \textit{Spitzer} survey \citep{megeath2012}. Chandra observations yielded a 95\% confidence upper limit of $6\times10^{-15}$ erg sec$^{-1}$cm$^{-2}$ \citep{audard2005}.

We observed VLBA 20 with the Submillimeter Array (SMA) on 2015 December 13 in the compact configuration at two frequencies (224.521 and 336.425 GHz) with an integration time of 400 min. Uranus and Callisto were used for flux calibration, 3C454.3 was used as the bandpass calibrator, and 0501-019 and 0607-085 were used as the gain calibrators. The data were reduced and calibrated following the standard prescriptions using the IDL package MIR as well as MIRIAD \citep{miriad}. A weak detection was obtained at 224.521 GHz with a measured flux of 2 $\pm$ 1 mJy. VLBA 20 was not detected at 336.425 GHz with 1$\sigma$ noise limit of 1 mJy. No molecular lines have been detected within the bandwidth of 4 GHz at either wavelength. 

The radio to near-infrared flux of VLBA 20 can be fitted by a power law with a slope of $\alpha=-0.72$, defined such that $S_\nu\propto\nu^\alpha$ (Figure \ref{fig:sed}). The spectrum is consistent with an AGN-related activity \citep[e.g.,][]{carilli1999}, and provides a clear indication that the radio emission has little or no thermal component.

\section{Multi-frequency VLBI observations}

\begin{deluxetable}{cccccc}
\tabletypesize{\scriptsize}
\tablewidth{0pt}
\tablecaption{Deconvolved gaussian model of VLBA 20. \label{tab:scatter}}
\tablehead{
\colhead{$\nu$\tablenotemark{a}} &\colhead{$S$} & \colhead{$a$} & \colhead{$b$} & \colhead{p.a.} & \colhead{q} \\
\colhead{(GHz)} & \colhead{(mJy)} & \colhead{(mas)} & \colhead{(mas)} & \colhead{(deg)} & \colhead{}
}
\startdata
5\tablenotemark{a} &  34.0$\pm$0.14 & 34.4$\pm$0.6 & 19.4$\pm$0.7 & 125.3$\pm$0.5 & 0.56$\pm$0.01\\
8.4 &  24.5$\pm$0.16 & 12.4$\pm$0.1 & 6.5$\pm$0.1 & 129.2$\pm$0.4 & 0.52$\pm$0.01\\
15.2 &  10.0$\pm$0.2 & 5.0$\pm$0.1 & 2.3$\pm$0.3 & 138.8$\pm$1.8 & 0.45$\pm$0.02\\
\enddata
\tablenotetext{a}{Multiple epochs are concatenated.}
\end{deluxetable}

In order to constrain the morphology of VLBA 20, and to complement existing 5 GHz observations from \citet{kounkel2017}, we obtained VLBA observations at 8.4 and 15.2 GHz. On both cases, the bandwidth was 256 MHz (8.304 -- 8.528 and 15.144 -- 15.368 GHz, respectively). The observations took place on 2017 July 1 and 2, with the total observing time of 8 hours shared between these dates. J0501-0159 was used to correct for the instrumental delays and J0529-0519 was used as the primary calibrator to estimate and remove residual rates, delays and phases. The observing strategy was to alternate between 150 s on source and 50 s on the primary calibrator for a combined total time of 152 min on source at 8.4 GHz for the two dates, and 120 s on source and 50 s on the primary calibrator for a combined total of 182 min on source at 15.2 GHz. The data were reduced in AIPS \citep{aips}; the full description of data processing is presented in \citet{ortiz-leon2017} and \citet{kounkel2017}.

\begin{figure}
 \centering 
        		\gridline{  
              \fig{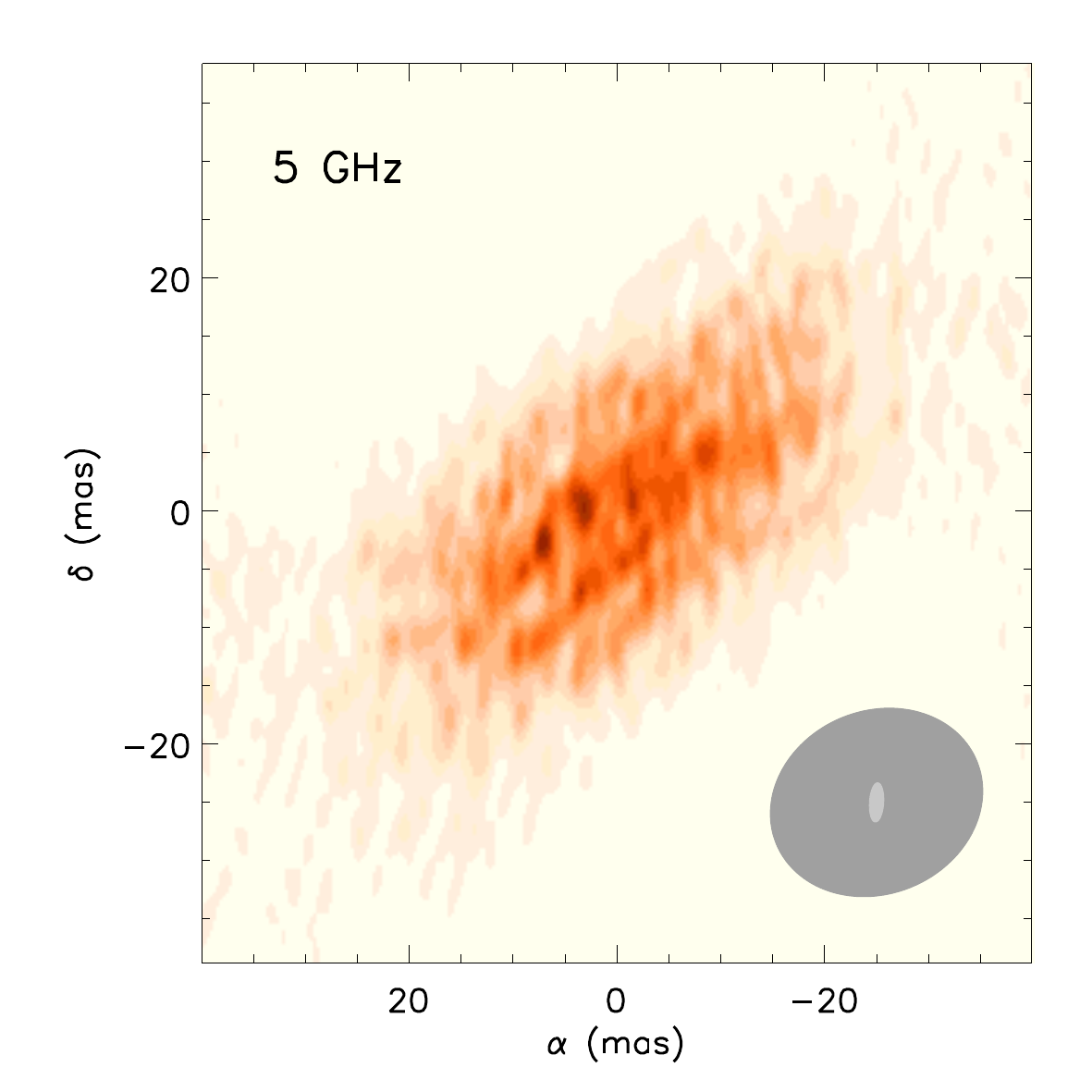}{0.25\textwidth}{}
              \hspace{-0.5cm}  
              \fig{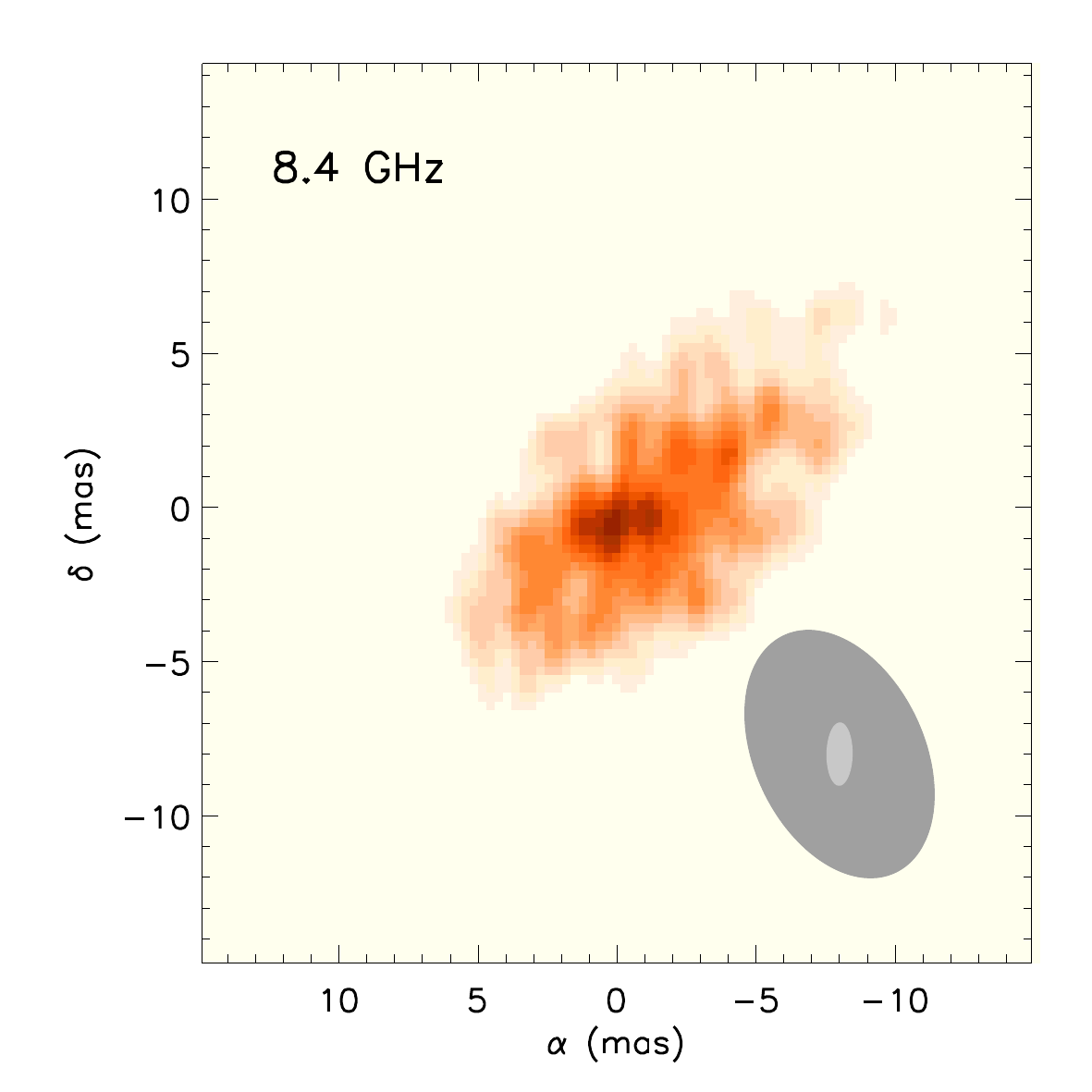}{0.25\textwidth}{}
        }\vspace{-1cm}
        		\gridline{  
              \fig{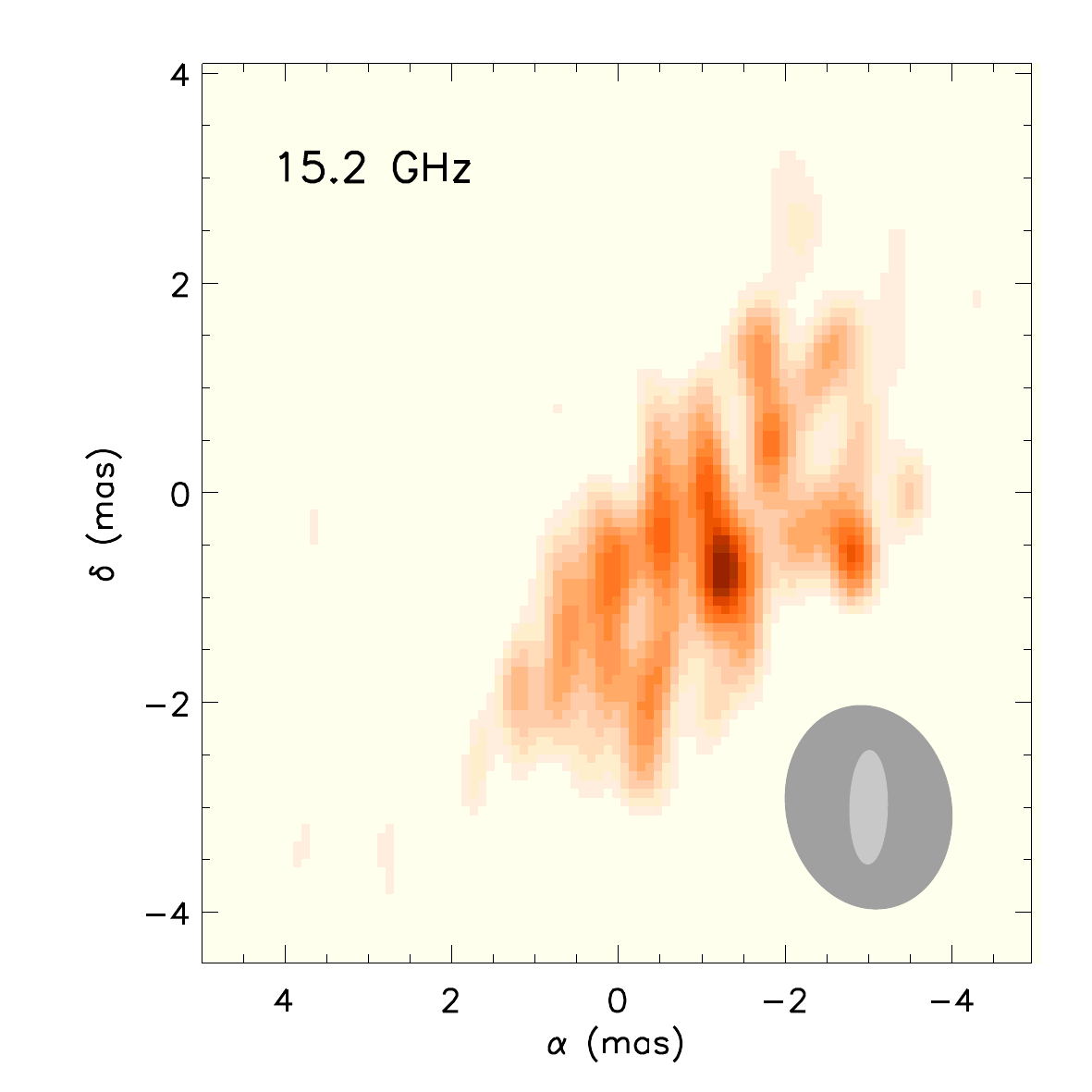}{0.25\textwidth}{}
        }
        \vspace{-0.8cm}
        
 \caption{VLBA 20 imaged at 5 GHz, 8.4 and 15.2 GHz, with the full UV range. Due to a lack of coherent signal at large baselines, imaging does not perfectly recover the smooth profile of the scattering disk, producing a stippling effect. Imaging is done using uniform weighting; other schemes do not substantially improve the image. Note that the size of the panel is different at each wavelength. Two beam sizes are shown, the beam corresponding to the full UV range (light gray), and the beam produced by limiting the UV range to the baselines with the coherent signal (dark gray). \label{fig:scatter2} }
\end{figure}

\begin{figure}
\epsscale{1}
\plotone{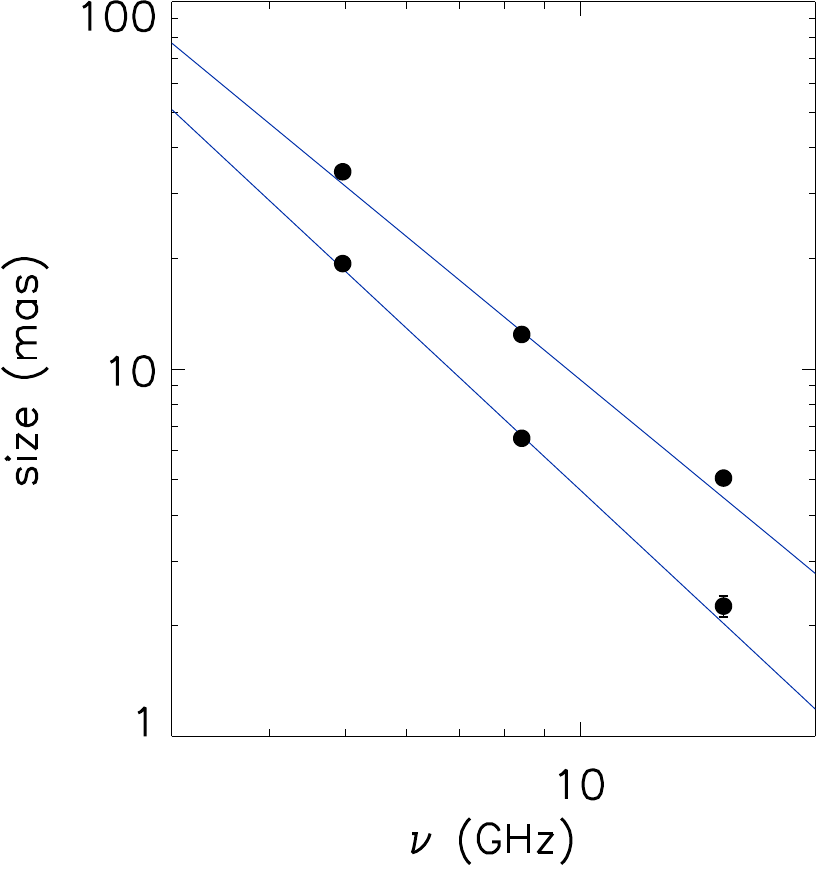}
\caption{Wavelength dependence of the semimajor and semiminor axes of VLBA 20, with the respective fitted slopes of -1.75$\pm$0.02 and -1.98$\pm$0.06. \label{fig:lambda}}
\end{figure}

The morphology observed at 5 GHz persists at the other wavelengths as well (Figure \ref{fig:scatter2}). We report on the dimensions of the source, which were obtained in the UV plane, by fitting a Gaussian profile using the CASA task uvmodelfit, in Table \ref{tab:scatter}. The dimensions are consistent with what is obtained in the image plane. At 8.4 GHz, no coherent signal was apparent beyond 23 M$\lambda$. Similarly, at 15.2 GHz, no coherent signal originated from beyond 45 M$\lambda$. The wavelength dependence in both the major and the minor axes of the scattering disk can be fitted by a power law with a slope $\alpha$ of -1.75$\pm$0.02 and -1.98$\pm$0.06 respectively (Figure \ref{fig:lambda}). For the minor axis, the relationship is consistent with $\alpha=-2$, which is the expected wavelength dependence for interstellar scattering. The major axis does deviate from this somewhat, but it is still largely consistent. It is possible that the deviation from the expected slope may be significant, although data at a larger number of frequencies would be required to more conclusively determine it. It may also be notable that both the semimajor and the semiminor axes at 15.2 GHz appear to be slightly above the fit, it is possible that at these high frequencies we may be starting to detect the intrinsic dimensions of the source.

It should be noted that at all the frequencies, the gain calibrator J0529-0519 does not appear to be resolved. In all observations its deconvolved components are generally consistent with a true point source. Strong signal is present in all the baselines. The same can be said of the secondary calibrators used in \citet{kounkel2017}, although J0539-0514, while also unresolved, does have its signal taper off on the longest baselines involving the Mauna Kea (MK) VLBA station in Hawaii. Other confirmed extragalactic sources towards the ONC might be weakly affected by scattering\footnote{On the other hand, members of the ONC are largely unaffected by angular broadening. Most of the photons would be scattered too strongly to converge at the position of the observer due to a wider range of the angles that reach the scattering screen, and, similarly to the other lensing-like effects, the remaining ones would not disperse over a sufficiently large solid angle to create as strongly scatter broadened image as a more distant source would be able to produce.
}, with the nominal deconvolved dimensions of up to $1-2$ mas at 5 GHz, but none as extreme as VLBA 20.

\begin{figure}
\epsscale{1}
\plotone{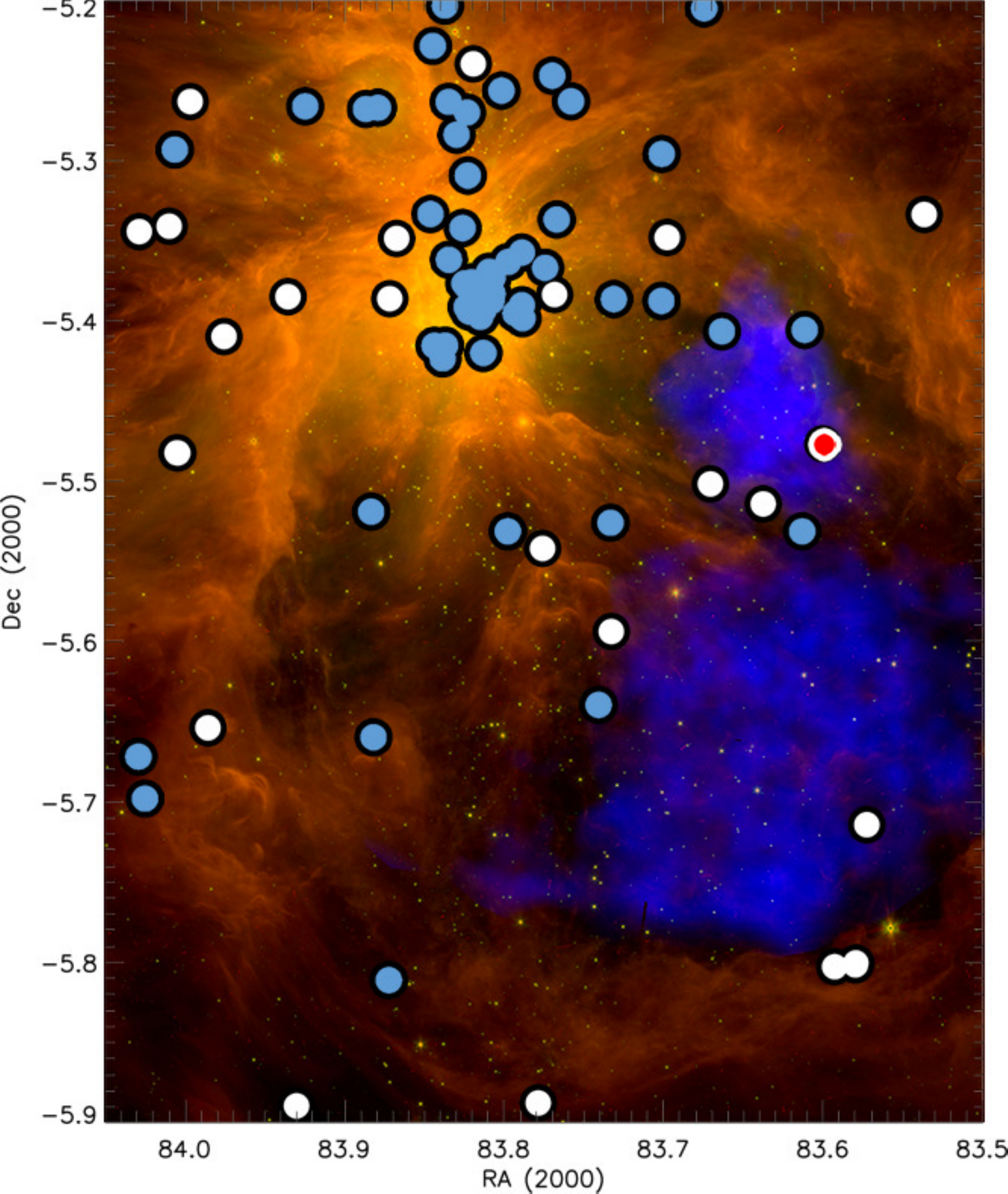}
\caption{Spitzer map of the ONC  with the superimposed diffuse x-ray emission (in blue) from \citet{gudel2008}. The dots are the observed radio sources with VLA from \citet{kounkel2014}. Blue - known members of the ONC. White - membership status unknown or likely extragalactic. Red - VLBA 20. \label{fig:map}}
\end{figure}

The origin of the scattering is most likely to be the hot X-ray emitting gas produced by the high velocity winds from $\theta^1$ Ori C, which was previously observed by \citet{gudel2008}. It is notable that very few extragalactic radio sources appear to be observed towards this gas\footnote{Another source is projected onto the diffuse X-ray emission, GBS-VLA J053418.01-054255.4. This source was not observed with the VLBA. It is not clear whether interstellar scattering would affect it or not as the X-ray emission towards it is patchy. In VLA observations it was point-like, and comparatively faint ($F_{4.5 GHz}=0.34$ mJy).} (Figure \ref{fig:map}), although a number are found near the outer bounds. In comparison to the other regions targeted by the VLA survey, a number of extragalactic sources near the HII region in the ONC may be underestimated by a factor of 1.5--2, although it is not clear what is responsible for it.

In general, the angular size $\theta_{\rm scatt}$ of a scattered point source depends both on the diffractive scale ($r_{\rm diff}$) of the scintillation and on the relative scattering geometry. Physically, the diffractive scale corresponds to the transverse length on the phase screen over which the random phase introduced by scattering shows a root-mean-squared difference of 1 radian. Thus, it quantifies the strength of the phase fluctuations of scattering. Specifically \citep[see, e.g.,][]{taylor1993,trotter1998}
\[
 \theta_{\rm scatt} \approx \frac{\sqrt{2 \ln{2}}}{\pi} \frac{\lambda}{(1+M) r_{\rm diff}} \approx 0.37 \frac{\lambda}{(1+M) r_{\rm diff}}\]
where $M$ is the ``magnification'' of the scattering, equal to the observer-scattering distance divided by the source-scattering distance. 

In the special case of an extragalactic source, $M \approx 0$. In this case, the angular broadening is independent of the distance from the observer to the scattering material: $\theta_{\rm scatt} \approx 0.37 \frac{\lambda}{r_{\rm diff}}$. Note that $\theta_{\rm scatt}$ scales approximately as $\lambda^2$, and $r_{\rm diff}$ scales approximately as $1/\lambda$. 

For VLBA 20, we find $\theta \approx 34\,{\rm mas}$ for the major axis at 5\,GHz, giving $r_{\rm diff} \approx 130~{\rm km}$ at that frequency. In contrast, typical angular broadening off the Galactic plane is ${\sim} 1\,{\rm mas} / \nu_{\rm GHz}^2$, giving $r_{\rm diff} \sim 10^5~{\rm km}$ at $5\,{\rm GHz}$ \citep{cordes2002}. Thus, the absolute density fluctuations along the line of sight to VLBA 20 (integrated through the scattering material) are stronger by a factor of ${\sim}1000$ than the `typical' values, although it is comparable in strength to the other regions that are associated with strong scattering.

\begin{figure}
\epsscale{1.1}
\plotone{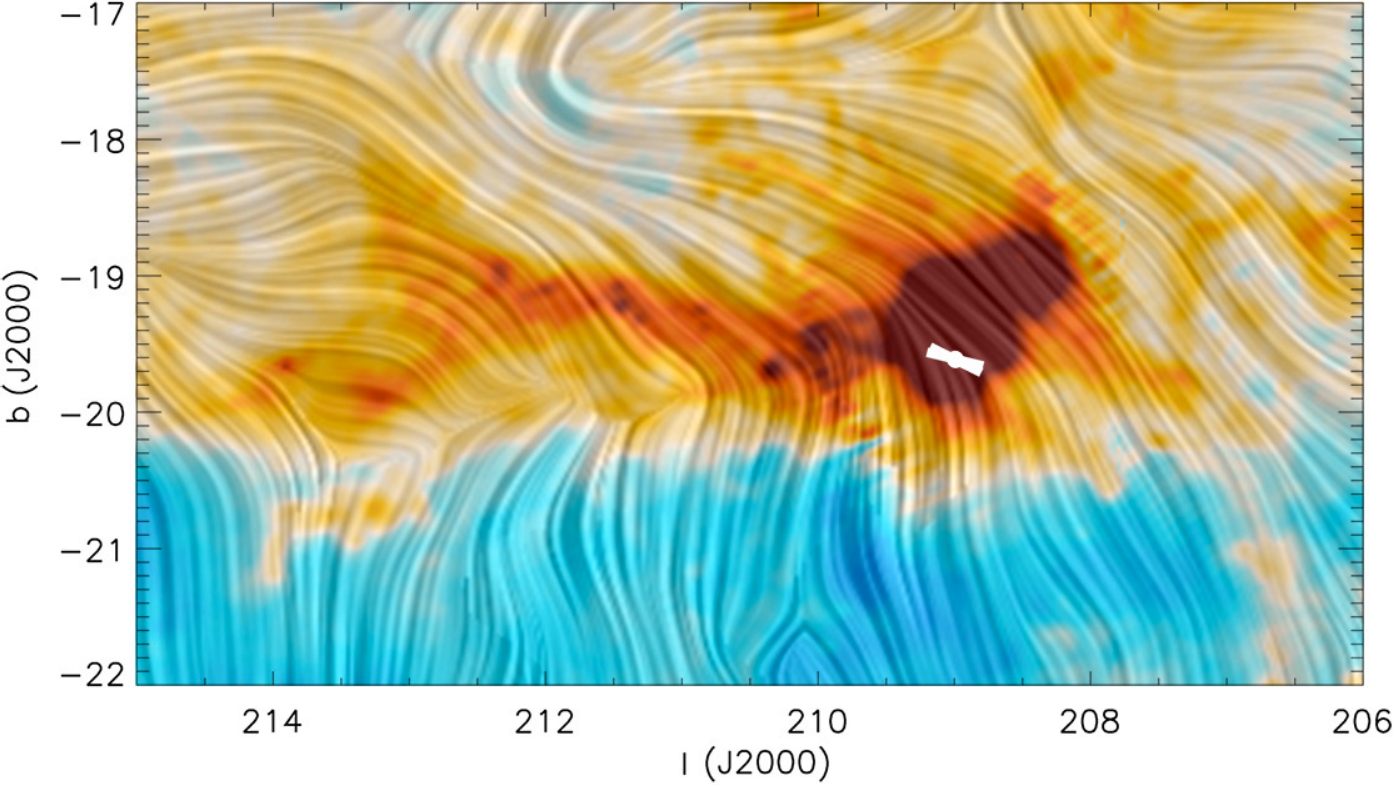}
\caption{Map of magnetic field lines towards the Orion from Planck \citep{planck-collaboration2016}, with the approximate position and orientation of VLBA 20. \label{fig:planck}}
\end{figure}

It is notable that the effect of interstellar scattering towards VLBA 20 is highly anisotropic in nature, with the axis ratio being consistent with 2:1 in all observations. Frequently, a higher axis ratio is apparent towards pulsars at frequencies less than 1 GHz \citep[e.g.,][]{rickett2006}. A comparable degree of anisotropy has been seen towards Sgr A* at 8.7 GHz, although the axis ratio appears to significantly decrease at 15.4 GHz and higher frequency data \citep{bower2014a,ortiz-leon2016}. VLBA 20 appears to have a scattering disk with a highest axis ratio at frequencies of 15.2 GHz, and possibly higher. If magnetic fields are an explanation for this anisotropy in this case, it suggests that they may be more coherent towards VLBA 20 than in the other HII regions associated with interstellar scattering. While the position angle does appear to slightly rotate with wavelength, the orientation of VLBA 20 is broadly consistent with being perpendicular to the magnetic field lines mapped by Planck, even though it does appear to deviate from it somewhat \citep{planck-collaboration2016}.  We note, however, that owing to the low angular resolution of the Planck data, it is difficult to make a more precise comparison, and higher resolution mapping of the magnetic fields is necessary to confirm it one way or another. (Figure \ref{fig:planck}).

At all frequencies the emission towards VLBA 20 does not have a smooth profile; rather, the scattering disk deconvolves into a series of stipples with the size comparable to the size of a beam (Figure \ref{fig:scatter2}). Efforts have been made in understanding whether this may be attributed to speckles, such as if the substructure is produced by the refractive noise that is present on long baselines \citep{johnson2015}. This has been found to be unlikely. Most probably, the stippling effect can be attributed to an imperfect reconstruction of a Gaussian profile while imaging due to the noise from the non-detection of the signal on long baselines. Removing those baselines does produce a smoother image (e.g., Figure \ref{fig:scatter1}). At higher frequencies the size of the degraded beam becomes comparable to the size of the source. Nonetheless, the anisotropy and the true dimensions of the source can be recovered during deconvolution even in the degraded images. 

\section{Conclusions}

We present VLBA observations of a peculiar radio source, VLBA 20. This source is projected near the ONC, but it is extragalactic, with non-thermal emission dominating the spectrum down to infrared wavelengths. VLBA 20 is heavily scattered by the HII region produced by the most massive stars of the ONC, the degree of scattering is comparable to what has been previously observed towards Cyg X3 \citep{wilkinson1994} and Sgr A* \citep{bower2014a}, although it is weaker than towards NGC 6334B \citep{trotter1998}. This source is highly anisotropic, with the scattering disk resolved to $\sim34\times19$ mas at 5 GHz. Higher frequency observations preserve the axis ratio, scaling the size by $\sim\nu^{-2}$. This anisotropy is most likely driven by the strong magnetic fields within the turbulent plasma.

\software{AIPS \citep{aips}, MIRIAD \citep{miriad}, MIR (https://www.cfa.harvard.edu/~cqi/mircook.html) }

\acknowledgments
L.L. and L.F.R. acknowledge the financial support of DGAPA, UNAM (project IN112417), and CONACyT, M\'{e}xico. The Long Baseline Observatory is a facility of the National Science Foundation operated under cooperative agreement by Associated Universities, Inc. The National Radio Astronomy Observatory is a facility of the National Science Foundation operated under cooperative agreement by Associated Universities, Inc. The Submillimeter Array is a joint project between the Smithsonian Astrophysical Observatory and the Academia Sinica Institute of Astronomy and Astrophysics and is funded by the Smithsonian Institution and the Academia Sinica. This work made use of the Swinburne University of Technology software correlator, developed as part of the Australian Major National Research Facilities Programme and operated under license \citep{deller2011}.

\bibliographystyle{aasjournal.bst}
\bibliography{scattering2}

\end{document}